\documentclass[a4paper, conference,10pt]{IEEEtran}
\usepackage{amsfonts}
\usepackage{mathrsfs}
\usepackage{amssymb,amsmath}
\usepackage{balance}

\usepackage{algorithm}
\usepackage{algorithmic}
\usepackage{amsmath,amssymb,epsfig,subfigure}
\usepackage{graphicx}
\usepackage{epstopdf}
\usepackage{theorem}
\usepackage{array,color}
\usepackage[compress]{cite}
\usepackage{enumerate}
\usepackage{bm}
\usepackage{balance}
\usepackage[symbol]{footmisc}
\usepackage[left=1in, right=1in, top=1in, bottom=1in]{geometry}

\theoremheaderfont{\normalfont\bfseries}

\usepackage{subfiles}
\graphicspath{{fig/}{../fig/}}

\begin{document}

\title{Deep Learning-Based Decoding for Constrained Sequence Codes}

\author{Congzhe~Cao$^{\dagger*}$, Duanshun Li$^{\ddagger*}$,
        and~Ivan~Fair$^\dagger$
\\
$^\dagger$Department of Electrical and Computer Engineering, University of Alberta, Edmonton, AB, Canada\\
$^\ddagger$Department of Civil and Environmental Engineering, University of Alberta, Edmonton, AB, Canada\\
Email:\{congzhe, duanshun, ivan.fair\}@ualberta.ca

}

\maketitle

\begin{abstract}
Constrained sequence codes have been widely used in modern communication and data storage systems. Sequences encoded with constrained sequence codes satisfy constraints imposed by the physical channel, hence enabling efficient and reliable transmission of coded symbols. Traditional encoding and decoding of constrained sequence codes rely on table look-up, which is prone to errors that occur during transmission. In this paper, we introduce constrained sequence decoding based on deep learning. With multiple layer perception (MLP) networks and convolutional neural networks (CNNs), we are able to achieve low bit error rates that are close to maximum a posteriori probability (MAP) decoding as well as improve the system throughput. Moreover, implementation of capacity-achieving fixed-length codes, where the complexity is prohibitively high with table look-up decoding, becomes practical with deep learning-based decoding. 

\end{abstract}
\footnotetext[1]{Congzhe Cao and Duanshun Li contributed equally to this work. 
}

	\section{Introduction}
	Constrained sequence (CS) codes have been widely used in communication and data storage systems to provide high transmission reliability \cite{Textbook}. Since the initial study of CS coding in Shannon's 1948 paper \cite{Shannon}, researchers have continuously worked in this area to design efficient CS codes that achieve code rates close to capacity with low implementation complexity \cite{Textbook, Franaszek, 8B10B_Ref,DC-freeRLL_Immink,FixedLength2,ImminkVLCodes,Motwani, BalancedModulation,FixedLength3,Pearson1,SystematicPearsonCoding,MyMinimalSet,MyJSAC,myPearson,My_MinimalSet,DNAStorageImmink}. Table look-up is the most widely used approach for encoding and decoding a fixed-length code that maps length-$k$ source words to length-$n$ codewords. Although many good codes have been proposed and used in practical systems, CS codes often suffer from the following drawbacks: i) normally advantage is not taken of whatever error control capability may be inherent in CS codes, therefore they are prone to errors that occur during transmission. ii) The capacity of most constraints is irrational, therefore it is difficult to construct a CS codebook with rate $k/n$ that is close to capacity without using very large values of $k$ and $n$. However, with large $k$ and $n$ values the time and implementation complexity of table look-up become prohibitive since a total of $2^k$ codewords exist in the codebooks of binary CS codes. Therefore, design of practical capacity-achieving fixed-length CS codes has been a challenge for many years. 
	
	\begin{figure*}[htbp]
		\begin{center}
			\includegraphics[width=12.5cm]{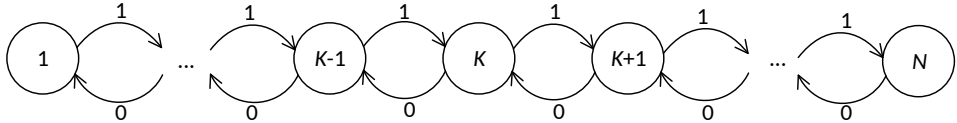}
			\makeatletter\def\@captype{figure}\makeatother
			\caption{FSM of DC-free constraints}\label{general}
		\end{center}
	\end{figure*}

	With the advancement of greater computational power and increasingly sophisticated algorithms, reinforcement learning (RL) has demonstrated impressive performance on tasks such as playing video games \cite{RL1} and Go \cite{RL2}. RL commonly uses Q-learning for policy updates in order to obtain an optimal policy that maps the state space to the action space \cite{sutton}, however, obtaining the update rule from table look-up, as traditionally has been done, becomes impossible with large state-action space. The invention of deep Q-networks that use deep neural networks (DNNs) to approximate the Q-function enables sophisticated mapping between the input and output, with great success \cite{deepRL}. Motivated by this approach, we hypothesized that it would be promising to replace look-up tables in CS codes with DNNs. Therefore, we propose using DNNs for CS decoding to overcome the drawbacks outlined above.
	
	Recently several works have reported the application of DNNs to the decoding of error control codes (ECCs) \cite{deeplearningECC1, deeplearningECC2, deeplearningECC3, deeplearningECC4, deeplearningECC5}. DNN enables low-latency decoding since it enables \emph{one-shot} decoding, where the DNN finds its estimate by passing	each layer only once \cite{deeplearningECC1, deeplearningECC3, deeplearningECC4}. In addition, DNNs can efficiently execute in parallel and be implemented with low-precision data types on a graphical processing unit (GPU), field programmable gate array (FPGA), or application specific integrated circuit (ASIC) \cite{deeplearningECC1, deeplearningECC3, deeplearningECC4, deeplearningECC5, deeplearningIntroduction}. It has been shown that, with short-to-medium length codewords, DNN-based decoding can achieve competitive bit error rate (BER) performance. However, since the number of candidate codewords becomes extremely large with medium-to-large codeword lengths, direct application of DNNs to ECC decoding becomes difficult because of the explosive number of layers and weights. In \cite{deeplearningECC4}, DNNs were employed on sub-blocks of the decoder, which were then connected via belief propagation decoding to enable scaling of deep learning-based ECC decoding. In \cite{deeplearningECC5}, the authors proposed recurrent neural network (RNN)-based decoding for linear codes, which outperforms the standard belief propagation (BP) decoding, and significantly reduces the number of parameters compared to BP feed-forward neural networks.
	
	To the best of our knowledge, no other work has yet been reported that explores deep learning-based decoding for CS codes. As we will show in the rest of our paper, deep learning fits well with CS decoding, naturally avoiding the explosive number of layers and weights that occur in ECC decoding. Throughout this paper we focus on the 4B6B CS code that has been employed in visible light communications \cite{VLC1}, however, we note that our discussion applies to any fixed-length CS code. The contributions of this paper are as follows:
	\begin{itemize}
		\item We explore multiple layer perception (MLP) networks and convolutional neural networks (CNNs) for CS decoding, and show that use of a CNN significantly reduces the complexity of decoding by employing the constraints that are inherent in CS codewords.
		\item We show that well-trained networks achieve BER performance that is very close to maximum a posteriori probability (MAP) decoding of CS codes, therefore increasing the reliability of transmission.
		\item We demonstrate that the design and implementation of fixed-length capacity-achieving CS codes with long codewords, which has long been impractical, becomes practical with deep learning-based CS decoding.
	\end{itemize}

	\section{Preliminaries}
	\subsection{CS codes}\label{CS_codes}

	CS encoders convert source bits into coded sequences that satisfy certain constraints imposed by the physical channel. Some of the most widely-recognized constraints include the RLL constraints that bound the number of encoded bits between consecutive transitions, and the DC-free constraints that bound the running digital sum (RDS) value of the encoded sequence given that RDS is the accumulation of encoded bit weights in a sequence, where a logic one is represented by weight $+1$ and a logic zero is represented by weight $-1$ \cite{Textbook}. Some other types of constraints include the Pearson constraint and the constraints that mitigate inter-cell interference in flash memories \cite{Pearson1, SystematicPearsonCoding,MyMinimalSet,MyJSAC,myPearson,My_MinimalSet}.
	
	CS encoders can be described by FSMs consisting of states, edges and labels. For example, the FSM of a DC-free constraint with an RDS value of $N$ is shown in Fig. \ref{general}, where the RDS can take any one of $N$ possible values. The \emph{capacity} of a constrained sequence $C$ is defined as \cite{Shannon}
	\begin{equation}\label{eq1}
	C=\lim_{m \rightarrow \infty}\frac{\log_2N(m)}{m}
	\end{equation}
	where $N(m)$ denotes the number of constraint-satisfying sequences of length $m$. Based on the FSM description and the adjacency matrix $\bf{D}$ \cite{Textbook}, we can evaluate the capacity of a constraint by calculating the logarithm of $\lambda_{max}$ which is the largest real root of the determinant equation \cite{Shannon}
	\begin{equation}\label{eq2}
	\det[{\bf{D}} - z{\bf{I}}] = {\bf{0}}
	\end{equation}
	where $\bf{I}$ is an identity matrix. The capacity is given as \cite{Shannon}
	
	\begin{equation}\label{eq3}
	C=\log_2\lambda_{max}
	\end{equation}
	with units bits of information per symbol.
	
	\subsection{4B6B code in visible light communications}
	
	Visible light communication (VLC), which refers to short-range optical wireless communication using the visible light spectrum from 380 nm to 780 nm, has gained much attention recently \cite{VLC1}. The simplest VLC relies on on-off keying (OOK) modulation, which is realized with DC-free codes to generate a constant dimming level of $50\%$. Three types of DC-free codes have been used in VLC standards to adjust dimming control and reduce flicker perception: the Manchester code, the 4B6B code and the 8B10B code \cite{VLC1}. We use the 4B6B code as a running example throughout this paper.
	
	The 4B6B code satisfies the DC-free constraint with $N=5$, which has a capacity of 0.7925 \cite{Textbook}. The codebook has 16 source words as shown in Table \ref{4B6B} \cite{VLC1}. Each source word has a length of 4 and is mapped to a codeword of length 6, which results in a code rate $R$ of $2/3$, and therefore an efficiency $\eta = R/C$ of $84.12\%$.
	
	\begin{table}[htbp]
		\centering
		\caption{The codebook of the 4B6B DC-free VLC code: $R=2/3, \eta=84.12\% $}\label{4B6B}
		\begin{tabular}{|c|c||c|c|}
			\hline
			Source word&  Codeword & Source word&  Codeword   \\
			\hline
			0000    &    001110   &  1000    &     011001  \\
			\hline
			0001   &     001101   &  1001   &     011010  \\
			\hline
			0010    &     010011  &  1010& 011100  \\
			\hline
			0011    &     010110  &  1011    &    110001  \\
			\hline
			0100    &     010101 &  1100    &    110010 \\
			\hline
			0101    &     100011 &  1101    &    101001 \\
			\hline
			0110    &     100110 &  1110    &    101010 \\
			\hline
			0111    &     100101 &  1111    &    101100 \\
			\hline
		\end{tabular}
	\end{table}

	\section{Deep learning-based CS decoding}
	\subsection{System model}
	
	\begin{figure*}[htbp]
		\begin{center}
			\includegraphics[width=130mm]{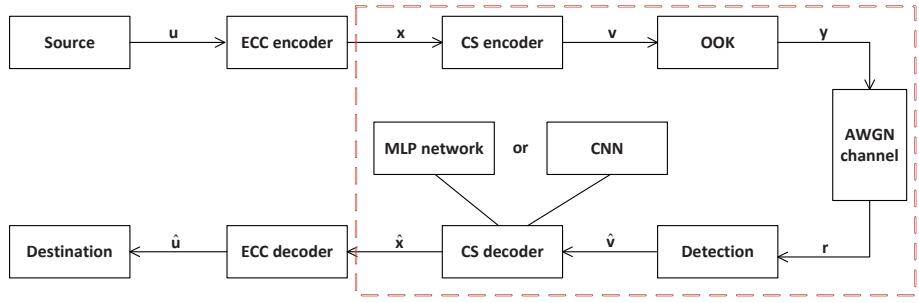}
			\makeatletter\def\@captype{figure}\makeatother
			\caption{System model}\label{system_model}
		\end{center}
	\end{figure*}
	
	A typical VLC system is shown in Fig. \ref{system_model}. Source bits $\bf{u}$ are encoded by an ECC encoder and a CS encoder (4B6B encoder) to generate coded bits $\bf{x}$ and $\bf{v}$, respectively. The coded bits are then modulated with OOK and transmitted via an additive white Gaussian noise (AWGN) channel. The received bits are
	\begin{equation}
	{\bf{r}} = {\bf{y}} + {\bf{n}}
	\end{equation}
	where ${\bf{n}}$ is the noise vector where each element is a Gaussian random variable with a zero mean and a variance of $\sigma^2$. The detector outputs symbol estimates $\bf{\hat{v}}$, and this sequence of estimates is decoded with the CS decoder (4B6B decoder) and ECC decoder successively to generate the estimate $\bf{\hat{u}}$. 
	In this paper we consider the framed components, and focus on the CS decoder that outputs $\bf{\hat{x}}$ as close as possible to $\bf{x}$. Throughout the paper we denote $|\bf{x}|$ as the size of vector $\bf{x}$.
	
	\subsection{MLP networks and CNNs}
	The fundamentals of deep learning are comprehensively described in \cite{deeplearningBook}. We employ both MLP networks and CNNs for CS decoding to predict $\bf{\hat{x}}$ given the input $\bf{\hat{v}}$. An MLP network has $L$ feed-forward layers. For each of the neurons in the MLP network, the output $y$ is determined by the input vector $\bf{t}$, the weights vector ${\bm{\theta}}$ and the activation function $g()$:
	\begin{equation}
	y = g({\boldsymbol{\theta}}\bf{t})
	\end{equation}
	where for the activation function we use the sigmoid function $g(z) = \frac{1}{1 + exp(-z)}$ and the rectified linear unit (ReLU) function $g(z) = max\{0,z\}$. A deep MLP network consists of many layers; the $i$th layer performs the mapping ${\bf{f}}^{(i)}:\mathbb{R}^{t_i} \rightarrow \mathbb{R}^{m_i}$, where $t_i$ and $m_i$ are the lengths of the input vector and the output vector of that layer, respectively. The MLP network is represented by:
	\begin{equation}
	{\bf{\hat{x}}} = {\bf{f}}^{(L)}({\bf{f}}^{(L-1)}(...{\bf{f}}^{(2)}({\bf{f}}^{(1)}(\bf{\hat{v}}))))
	\end{equation}
	
	The use of CNNs has recently achieved impressive performance in visual recognition, classification and segmentation, etc \cite{deeplearningBook}. It employs convolutional layers to explore local features instead of extracting global features with fully connected layers as in MLP networks, thus greatly reducing the number of weights that need to be trained and making it possible for the network to grow deeper \cite{resnet}. Different from visual tasks where the input colored images are represented by three dimensional vectors, the input vector $\bf{\hat{v}}$ in our task of CS decoding is a one dimensional vector. For a convolutional layer with $F$ \emph{kernels} given as: ${\bf{q}}^{f} \in \mathbb{R}^{1 \times |{\bf{q}}|}, f = 1,....,F$, the generated \emph{feature map} ${\bf{\hat{p}}}^f \in \mathbb{R}^{1 \times |{\bf{\hat{p}}}|}$ from the input vector ${\bf{\hat{v}}} \in \mathbb{R}^{1 \times |{\bf{\hat{v}}}|}$ satisfies the following dot product:
	\begin{equation}
	{p}^{f}_i = \sum_{l=0}^{|{\bf{q}}|-1}q^f_l {\hat{v}}_{1+s(i-1)+l}
	\end{equation}
	where $s \ge 1$ is the \emph{stride}. Usually convolutional layers are followed by \emph{pooling layers} such that high-level features can be extracted at the top layers. However, as we will show, pooling may not fit in CS decoding and therefore our implementation of CNN does not include pooling layers.
	
	\subsection{Training method}
	
	In order to keep the training set small, we follow the training method in \cite{deeplearningECC1} where the DNN was extended with additional layers of modulation, noise addition and detection that have no additional parameters that need to be trained. Therefore, it is sufficient to work only with the sets of all possible noiseless codewords ${\bf{v}} \in \mathbb{F}_2^{|\bf{v}|}, \mathbb{F}_2 \in \{0,1\}$, i.e., training epoches, as input to the DNNs. For the additional layer of detection, we calculate the log-likelihood ratio (LLR) of each received bit and forward it to the DNN. 
	We use the mean squared error (MSE) as the loss function, which is defined as:
	\begin{equation}
	L_{MSE} = \frac{1}{|{\bf{u}}|} \sum_i(u_i-\hat{u}_i)^2.
	\end{equation}
	
	Both the MLP networks and CNNs employ three hidden layers. The detailed parameters are discussed in the next section. We aim at training a network that is able to generalize, i.e., we train at a particular signal-to-noise ratio (SNR), and test it within a wide range of SNRs. The criterion for model selection that we employ follows \cite{deeplearningECC1}, which is the normalized validation error (NVE) defined as:
	\begin{equation}
	NVE(\rho_t) = \frac{1}{S}\sum_{s=1}^S \frac{BER_{DNN}(\rho_t,\rho_{v,s})}{BER_{MAP}(\rho_{v,s})},
	\end{equation}
	where $\rho_{v,s}$ denotes the $S$ different test SNRs. $BER_{DNN}(\rho_t,\rho_{v,s})$ denotes the BER achieved by the DNN trained at SNR $\rho_t$ and tested at SNR $\rho_{v,s}$, and $BER_{MAP}(\rho_{v,s})$ denotes the BER of MAP decoding of CS codes at SNR $\rho_{v,s}$. The networks are trained with sufficient epoches until the loss $L_{MSE}$ converges.
	
	\begin{table*}[htbp]
		\centering
		\caption{Parameters of the MLP networks and CNNs trained for CS decoding with different frames}\label{parameters}
		\begin{tabular}{|c||c|c||c|c|}
			\hline
			\# of frame&  \# of neurons (MLP) & \# of parameters (MLP) &  \# of kernels (CNN) & \# of parameters (CNN)   \\
			\hline
			1    &  [32,16,8]    &   924   &  [8,12,8]    & 760\\
			\hline
			2   &     [64,32,16]    & 3576    &  [8,14,8]    & 1374\\
			\hline
			3    &     [128,64,32]  &  13164 & [8,16,8] & 2372\\
			\hline
			4    &     [128,128,64]   &  29008    &    [16,16,12]  & 5676\\
			\hline
			5    &     [256,128,64]  &   50338   &    [16,32,12] & 9536\\
			\hline
		\end{tabular}
	\end{table*}
	
	\begin{table*}[htbp]
		\centering
		\caption{Structures of the CNNs for CS decoding}\label{cnnparameters}
		\begin{tabular}{|c|c|c|c|}
			\hline
			layer  &  kernal size / stride & input size   & padding \\
			\hline
			OOK    &  N/A    &  $1 \times |{\bf{v}}|$  &  N/A\\
			\hline
			Adding noise   &   N/A      &  $1 \times |{\bf{v}}|$ &  N/A\\
			\hline
			LLR    &  N/A   & $1 \times |{\bf{v}}|$ &  N/A\\
			\hline
			Convolution   &   $1 \times 3$ / 1     & $1 \times |{\bf{v}}|$    &  no    \\
			\hline
			Convolution    &  $1 \times 3$ / 1    &  $1 \times (|{\bf{v}}|-2) \times h_1$    &  yes  \\
			\hline
			Convolution    &  $1 \times 3$ / 1    &   $1 \times (|{\bf{v}}|-2) \times h_2$   &  yes  \\
			\hline
			Fully connected    &   N/A    &  $1 \times ((|{\bf{v}}|-2) \times h_3)$      &  N/A\\
			\hline
			Sigmoid    &  N/A     &  $1 \times (|{\bf{x}}|)$    &  N/A  \\
			\hline
		\end{tabular}
	\end{table*}

	\section{Results and outlook}
	We use the notation ${\bf{h}} = [h_1, h_2,....,h_L]$ to represent a network with $L$ hidden layers, where we denote the number of neurons in a fully connected layer, or the number of kernels in a convolutional layer, as $h_i$ in layer $i$. In recent works that apply DNNs to decode ECCs, the training set explodes rapidly as the source word length grows. For example, with a rate 0.5 $(n=1024, k=512)$ ECC, one epoch consists of $2^{512}$ possibilities of codewords of length 1024, which results in very large complexity and makes it difficult to train and implement DNN-based decoding in practical systems \cite{deeplearningECC1, deeplearningECC2, deeplearningECC3, deeplearningECC4}. However, we note that in CS decoding, this problem does not exist since CS source words are typically considerably shorter, possibly up to a few dozen symbols \cite{Textbook, Franaszek, 8B10B_Ref,DC-freeRLL_Immink,FixedLength2,ImminkVLCodes,Motwani, BalancedModulation,FixedLength3,Pearson1,SystematicPearsonCoding,MyMinimalSet,MyJSAC,myPearson,My_MinimalSet,DNAStorageImmink}. This property fits deep learning based-decoding well.
	\subsection{BER performance}\label{BER}
	\subsubsection{Frame-by-frame decoding}\label{1frame}
	First we consider frame-by-frame transmission, where the 4B6B codewords are transmitted and decoded one-by-one, i.e., $|{\bf{v}}|=6$. We will later consider processing multiple frames simultaneously to improve the system throughput. Note that in the VLC standard, two 4B6B look-up tables can be used simultaneously \cite{VLC1}.
	
	We compare performance of deep learning-based decoding with traditional table look-up decoding that generates hard-decision bits. That is, the traditional detector estimates the hard decision $\hat{\bf{v}}$, and the CS decoder attempts to map $\hat{\bf{v}}$ to a valid source word to generate $\hat{\bf{x}}$. If the decoder is not able to locate $\hat{\bf{v}}$ in the code table due to erroneous estimation at the detector, the decoder determines the codeword that is closest to $\hat{\bf{v}}$ in terms of Hamming distance, and then outputs the corresponding source word. We also implement the maximum likelihood (ML) decoding of CS codes, where the codeword with the closest Euclidean distance to the received noisy version of codeword is selected and the corresponding source word is decoded. We assume equiprobable 0s and 1s in source sequences, thus ML decoding is equivalent to MAP decoding of CS codes since each codeword has an equal occurrence probability.
	
	Table \ref{parameters} shows the parameters of the MLP networks and the CNNs for a variety of tasks. The DNNs are trained at an SNR of 1 dB, using \emph{Adam} for stochastic gradient descent optimization \cite{Adam}. With $|{\bf{v}}| = 6$, the MLP network we trained for frame-by-frame decoding has three hidden layers [32,16,8] and 924 trainable parameters. Its BER performance is shown in Fig. \ref{1frame}, which shows that DNN-based decoding achieves a BER that is very close to MAP decoding of CS codes, and outperforms conventional table look-up decoding by $\sim$2.2 dB.
	
	\begin{figure}[htbp]
		\begin{center}
			\includegraphics[width=\linewidth]{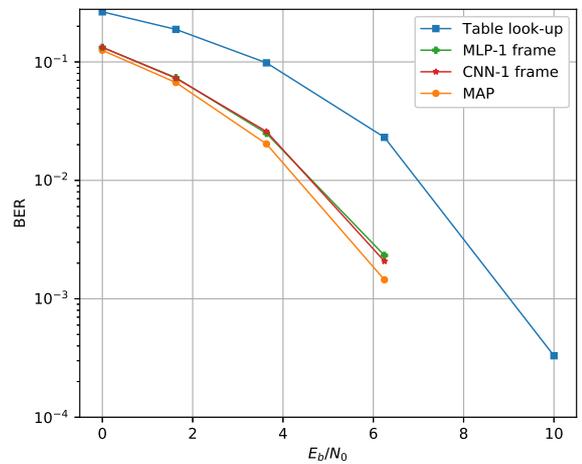}
			\makeatletter\def\@captype{figure}\makeatother
			\caption{comparison of BER performance with frame-by-frame transmission, i.e., $|{\bf{v}}|=6$}\label{1frame}
		\end{center}
	\end{figure}
	
	We then investigate employing CNNs for this task. With ECC decoding, CNNs and MLP networks have roughly the same complexity, i.e., a similar number of weights in order to achieve similar performance \cite{deeplearningECC3}. In what follows we outline our findings that are unique to CS decoding.
	
	In Table \ref{cnnparameters} we outline the structure of the CNNs we apply for CS decoding. ReLU is used as the activation function for each convolutional layer. We note that CS codes always have inherent constraints on their codewords such that they match the characteristics of the transmission or storage channel. For example, the 4B6B code in Table \ref{4B6B} always has an equal number of logic ones and logic zeros in each codeword, and that the runlength is limited to four in the coded sequence for flicker reduction. These \emph{low-level features} can be extracted to enable CNNs to efficiently learn the weights of the kernels, which results in significantly reduced complexity compared to MLP networks. For example, although similar BER performance is achieved by the [32, 16, 8] MLP network and the [6, 10, 6] CNN, the number of weights in the CNN is only $82\%$ of that in the MLP network. With larger networks the complexity reduction is more significant, as we will show in the next subsection.
	
    Another finding we observe during training of a CNN is that pooling layers which are essential component structures in CNNs for visual tasks, may not be required in our task. The reason is that in visual tasks, pooling is often used to extract \emph{high-level} features of images such as shapes, edges or corners. However, CS codes often possess low-level features only, and we find that adding pooling layers may not assist CS decoding. Therefore, no pooling layer is used in our CNNs, as indicated in Table \ref{cnnparameters}. Fig. \ref{1frame} shows that use of a CNN achieves similar performance to the use of an MLP network, and that it also approaches the performance of MAP decoding.
	
	\subsubsection{Improving the throughput}\label{nframe}
	We now consider processing multiple frames in one time slot in order to improve the system throughput. The system throughput can be enhanced by increasing the optical clock rate, which has its own physical limitations, or by processing multiple 4B6B codewords in parallel. The VLC standard allows two 4B6B codes to be processed simultaneously \cite{VLC1}. Now we show that DNNs can handle larger input size where $|{\bf{v}}|$ is a multiple of 6, thus system throughput can be enhanced by using one of those DNNs or even using multiple DNNs in parallel.
	
	\begin{figure}[htbp]
		\begin{center}
			\includegraphics[width=\linewidth]{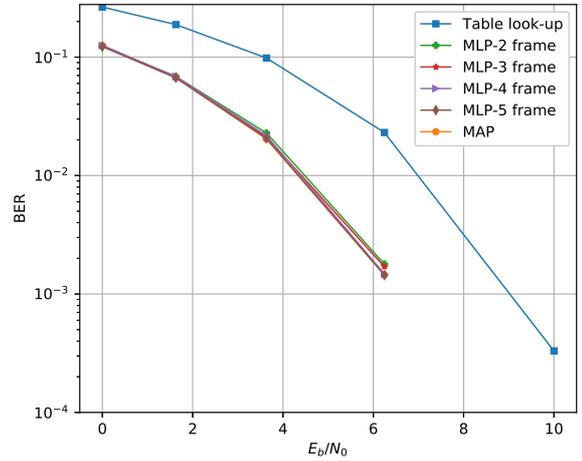}
			\makeatletter\def\@captype{figure}\makeatother
			\caption{BER performance of MLP networks with multiple frames processed simultaneously, i.e., $|{\bf{v}}|=12, 18, 24, 30$}\label{nframe_MLP}
		\end{center}
	\end{figure}
	
	\begin{figure}[htbp]
		\begin{center}
			\includegraphics[width=\linewidth]{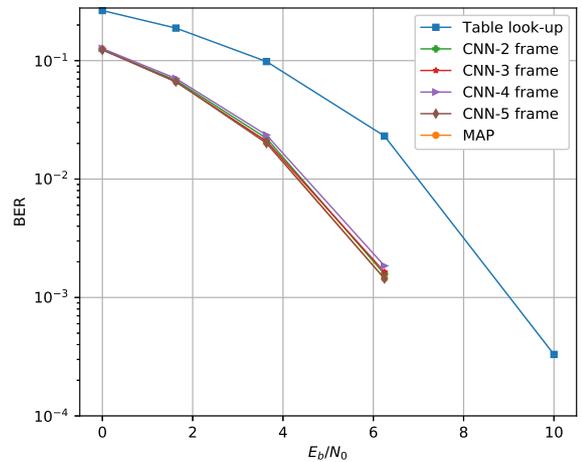}
			\makeatletter\def\@captype{figure}\makeatother
			\caption{BER performance of CNNs with multiple frames processed simultaneously, i.e., $|{\bf{v}}|=12, 18, 24, 30$}\label{nframe_CNN}
		\end{center}
	\end{figure}

	Figs. \ref{nframe_MLP} and \ref{nframe_CNN} present the BER performance of MLP networks and CNNs respectively, where the parameters of those networks are shown in Table \ref{parameters}. These figures demonstrate that both MLP networks and CNNs are able to achieve BERs very close to MAP decoding, while the CNNs have significantly lower complexity than the MLP networks for the reason outlined above. With larger $|{\bf{v}}|$, it becomes easier for the CNN to extract the low-level features from the longer input seqeuences and learn the weights, and thus the complexity reduction with CNN is more significant. For example, when processing five frames simultaneously, the CNN has less than $1/5$ of the parameters that need to be trained for the MLP network. Note that we consider the networks shown in Table \ref{parameters} to be small (ResNet \cite{resnet} has a few million parameters to train). Therefore, we anticipate that it will be practical to achieve further improvement on system throughput with larger networks.

	\subsection{Paving the way to capacity-achieving CS codes}\label{capacity-achieving}
		
	As we outlined in Section I, it has not been an easy task to design fixed-length capacity-achieving CS codes. As determined by equations \eqref{eq1}-\eqref{eq3}, the capacity of a constraint is most likely irrational, which can be approached with fixed-length codes of rate $R=k/n$ only with very large $k$ and $n$ values. This hinders implementation of table look-up encoding and decoding for capacity-achieving CS codes. For example, as we show in Section II-B, the 4B6B code achieves $84.12\%$ of the capacity of a DC-free code with 5 different RDS values, which has $C=0.7925$. In Table \ref{knvalues} we list, for increasing values of $k$, values of $n$ such that code rate $R$ approaches capacity. This table shows that with $k=11,15,19$, it could be possible to construct fixed-length codes with efficiencies that exceed $99\%$. With $k$ as large as $79$ and $n=100$, the code would have rate 0.79 and efficiency $99.68\%$. However, a table look-up codebook with $2^{k}$ source word-to-codeword mappings becomes impractical to implement as $k$ grows large. Other examples are the $k$-constrained codes recently developed for DNA-based storage systems in \cite{DNAStorageImmink}. Those fixed-length 4-ary $k$-constrained codes have rates very close to the capacity, however they require very large codebooks. For example, with method B in \cite{DNAStorageImmink}, for the $k$-constrained code with $k=1,2,3,4$, the codebooks have $4^{10},4^{38},4^{147},4^{580}$ source word-to-codeword mappings respectively.

\begin{table}
	\centering
	\caption{The highest code rate and efficiency with fixed-length CS codes for the DC-free constraint with five different RDS values, $C=0.7925$}\label{knvalues}
	\begin{tabular}{c|c|c|c||c|c|c|c}
		$k$&$n$&$R$&$\eta$&$k$&$n$&$R$&$\eta$ \\
		\hline
		1&2&0.5000&63.09\%&11&14&0.7857&99.14\%\\
		2&3&0.6667&84.12\%&12&16&0.7500&94.64\%\\
		3&4&0.7500&94.64\%&13&17&0.7647&96.49\%\\
		4&6&0.6667&84.12\%&14&18&0.7778&98.14\%\\
		5&7&0.7143&90.13\%&15&19&0.7895&99.62\%\\
		6&8&0.7500&94.64\%&16&21&0.7619&96.14\%\\
		7&9&0.7778&98.14\%&17&22&0.7727&97.51\%\\
		8&11&0.7273&91.77\%&18&23&0.7826&98.75\%\\
		9&12&0.7500&94.64\%&19&24&0.7917&99.89\%\\
		10&13&0.7692&97.06\%&20&26&0.7692&97.06\%\\
	\end{tabular}
\end{table}

	\begin{figure}[htbp]
	\begin{center}
		\includegraphics[width=\linewidth]{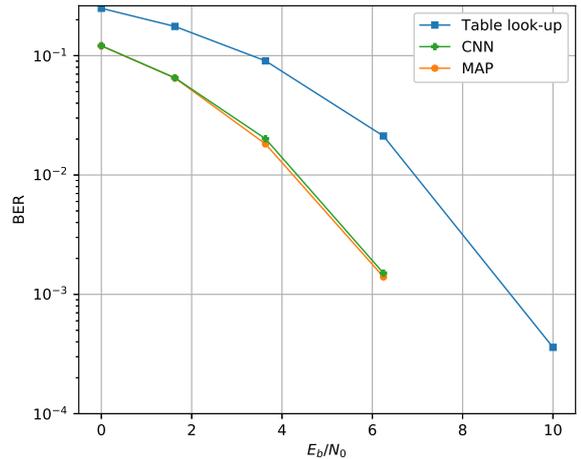}
		\makeatletter\def\@captype{figure}\makeatother
		\caption{BER performance of DNN-based decoding for a constrained sequence code with $2^{20}$ possibilities of mappings }\label{CNN_shuffle}
	\end{center}
\end{figure}

With DNNs, however, it becomes practical to handle a large set of source word-to-codeword mappings which has long been considered impractical with table look-up decoding. This paves the way for practical design and implementation of fixed-length capacity-achieving CS codes. Appropriate design of such codes is a practice of using standard algorithms from the rich theory of CS coding, such as Franaszek's recursive elimination algorithm \cite{Franaszek}, or the sliding-block algorithm \cite{ACH1, ACH2} with large $k$ and $n$ values to determine the codebooks. We propose implementing both the encoder and the decoder with DNNs. Although here we focus on decoding, similar to DNN-based decoders, CS encoders map noiseless source words to codewords, and can also be implemented with DNNs.

We now demonstrate that the proposed DNN-based decoders are able to map long received words from a CS code to their corresponding source words. We concatenate five 4B6B codebooks, where each 4B6B codebook is randomly shuffled in terms of its source word to codeword mappings, to generate a large codebook with $2^{20}$ entries of mappings. We train and test a CNN with ${\bf{h}} = [16,32,12]$ that has 9536 weights to perform decoding. From Fig. \ref{CNN_shuffle}, we can see that this CNN is capable of decoding the received noisy version of the large set of received words. The BER is close to MAP decoding, and outperforms the table look-up decoding approach that we implemented as a benchmark. Therefore, the design and implementation of DNN-based CS decoding is practical with long CS codewords because of their low-level features that we can take advantage of to simply decoding. 

We also note that DNNs that are proposed in communication systems could have many more parameters than the 9536 we use in this network. For example, \cite{SCMA} proposes an MLP network with 4 hidden layers where each layer has 512 neurons, which results in at least $512\times512\times3=786432$ parameters. Thus a larger CNN can be trained to decode fixed-length capacity-approaching CS codes with longer codewords.

	\section{conclusion}
    In this paper, we introduced deep learning-based CS decoding. We studied two types of DNNs, namely MLP networks and CNNs, and found that both networks can achieve BER performance close to MAP decoding as well as improve the system throughput, while CNNs have significantly lower complexity than MLP networks since they are able to efficiently exploit the inherent constraints imposed on the codewords. Furthermore, we have shown that the design and implementation of fixed-length capacity-achieving CS codes that has long been impractical, becomes practical with deep learning-based decoding. This new observation paves the way to deploying highly efficient CS codes in practical communication and data storage systems. 

	\balance

\end{document}